\documentclass[preprint,12pt]{elsarticle}

\usepackage{graphicx}
\usepackage{amsmath,amssymb,braket,bm}
\usepackage{booktabs,tabularx,multirow,array}
\usepackage{subcaption}
\usepackage{algorithm,algorithmic}
\usepackage{afterpage,float}
\usepackage[export]{adjustbox}
\usepackage{comment}

\begin{document}

\begin{frontmatter}



\title{Quantum Energy Teleportation across Multi-Qubit Systems using W-State Entanglement}

\author[inst1]{Alif Elham Khan\corref{cor1}}
\author[inst1]{Humayra Anjum}
\author[inst1]{Mahdy Rahman Chowdhury}
\cortext[cor1]{E‑mail: \texttt{alif.khan1@northsouth.edu}}

\affiliation[inst1]{%
  organization={Department of Electrical and Computer Engineering, North South University},
  city={Dhaka},
  postcode={1229},
  country={Bangladesh}
}

\begin{abstract}
Quantum-energy teleportation (QET) has so far only been realized on a
two-qubit platform.  Real-world communication, however, typically
involves multiple parties.  Here we design and experimentally
demonstrate the first multi-qubit QET protocol using a robust W-state
multipartite entanglement.  Three-, four-, and five-qubit circuits were
executed both on noiseless simulators and on IBM superconducting
hardware.  In every case a single sender injects an energy $E_{0}$ that is
then deterministically and decrementally harvested by several remote
receivers, confirming that energy introduced at one node can be
redistributed among many entangled subsystems at light-speed-limited
classical latency.  Our results open a practical route toward
energy-aware quantum networks.

\end{abstract}




\end{frontmatter}


\section{Introduction}
Quantum Energy Teleportation (QET) is a protocol that aims to achieve successful energy transportation via local operations and classical communication without breaking any known physical laws \cite{hotta2014quantum} \cite{hotta2011quantum}. Since its proposal in the year 2008, it has become the focus of studies and trials. The methodology entails using local operations that rely on data from a remote measurement to extract zero-point energy from entangled many-body systems \cite{hotta2009quantum}. More than a decade later, the first demonstration of QET has taken place on real quantum hardware \cite{ikeda2023demonstration} \cite{ikeda2023first}. The experiment, run on a two-qubit circuit, has surpassed the boundaries of large distances by transporting energy much faster than the time scale of heat generation in the natural time evolution. However, to truly leverage the power of long-distance quantum energy teleportation, we must investigate the feasibility of QET across multi-qubit systems because exchanging energy between multiple subsystems is more natural. In this study, we devised a quantum circuit employing W-state multipartite entanglement to observe energy transfer behavior across multiple entangled qubits. 
\begin{figure}[!htb]
    \centering
    \begin{subfigure}{0.65\textwidth}
        \centering
        \includegraphics[width=\columnwidth,valign=t]{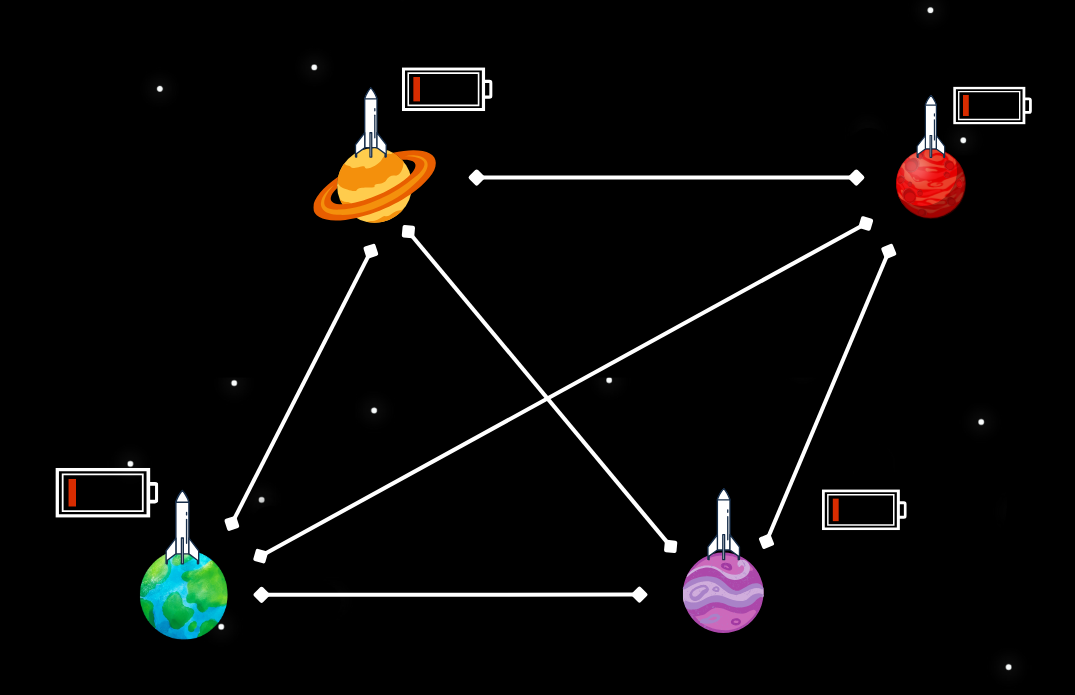}
        \caption{Before injecting energy}
        \label{fig1:subfigure-a}
    \end{subfigure}
  
    \begin{subfigure}{0.65\textwidth}
        \centering
       \includegraphics[width=\columnwidth,valign=t]{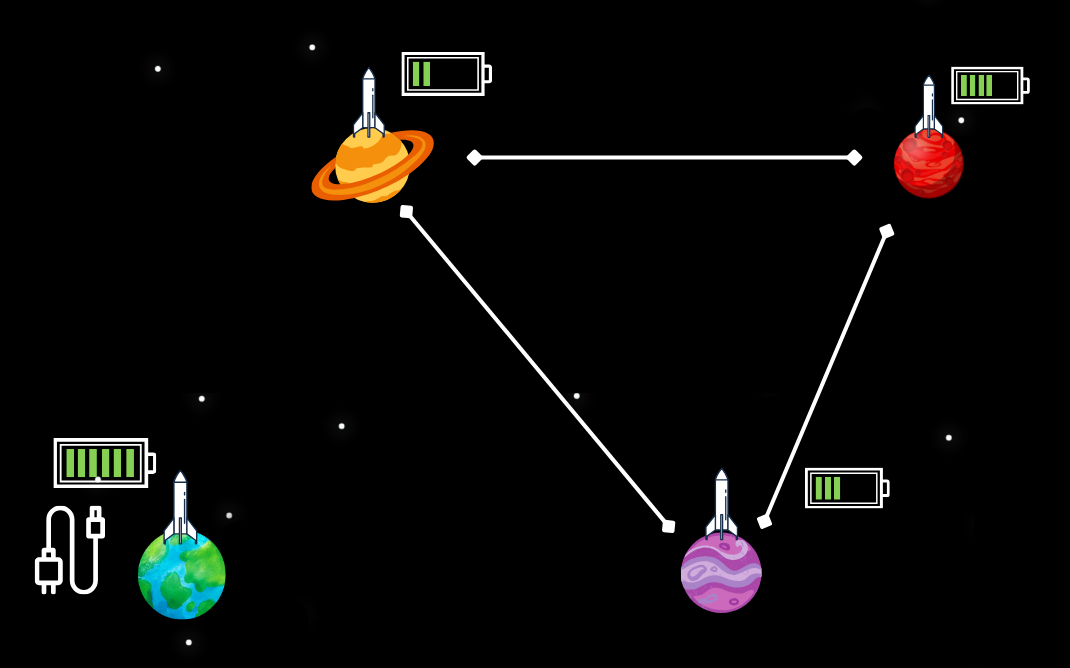}
     \caption{After injecting energy}
        \label{fig1:subfigure-b}
    \end{subfigure}
    
    \caption{(a) Entanglement exists between a sender and three receivers, but energy has not yet been injected. (b) Energy injected into the sender’s system is instantly distributed to the receivers located at far-off locations using Multi-Qubit Quantum Energy Teleportation. } 
     \label{engDist}
\end{figure}
It is essential to understand that QET does not promise to create energy out of a vacuum. Instead, energy can be unlocked using knowledge from a sender with energy in a far-off location. From this perspective, QET looks less like energy creation and more like the teleportation of energy from one place to another \cite{bennett1993teleporting}. In line with these conceptual foundations, we have devised a quantum circuit where energy is first injected into one qubit, Alice, and later harvested from the other entangled qubits: Bob, Charlie, Duke, Eric, etc, using the knowledge obtained from Alice. The multipartite entanglement exists in such a way that all of the remaining qubits are entangled to Alice such that when energy is injected into Alice, others can harvest it from far-off locations with the information provided by Alice. 
All of the participants are entangled using W-state entanglement \cite{wang2008preparation}. So, it is possible to inject energy at any node and harvest it from any other node. Injecting energy into the sender's system ensures that the receivers receive energy much faster than the time scale of heat generation in the natural time evolution. However, the QET protocols are based on local observation and classical communication (LOCC) \cite{chitambar2014everything}, which limits the protocols to the speed of light because the information on how to harvest the energy is sent through a classical channel bounded by the speed of light.

\section{Multiqubit System for QET}

QET uses quantum entanglement \cite{bub2001quantum}, a phenomenon where particles become correlated in such a way that the state of one particle is dependent on the state of another, regardless of the distance between them. In a QET scenario, one system (the sender) is entangled with another system (the receiver). The sender is provided with energy. This energy is teleported to the receiver through the entangled connection without physically traversing the space between them.

Quantum energy teleportation has only been tested in bipartite systems \cite{wu2004quantum} till now, leaving a broad scope of exploration for its behavior in multipartite systems. However, multipartite systems are complex. If one qubit is measured, then all of the entangled states will collapse. This is the case with Greenberger-Horne-Zeilinger (GHZ) state \cite{sk2021quantum}. If we measure an arbitrary qubit from a multi-qubit system, the system collapses into a single state, and entanglement ceases to exist. Measuring any one of the qubits, therefore, destroys the multipartite entanglement.

However, a successful energy exchange among multiple parties entails that the entanglement is preserved between parties that have not yet been measured. In contrast to the GHZ state, in the W state \cite{dur2000three}, all of the qubits are entangled, and the state is highly robust against single-particle measurements, meaning that the entanglement is preserved even if one of the qubits is lost or measured \cite{zang2015generating}. This serves our goal of preserving the state of the remaining subsystem when one of the particles is measured for harvesting energy. So, we created our circuit using W-state entanglement such that all the particles involved are entangled. For better visualization, figure \ref{fig1:subfigure-a} shows that four systems located at arbitrary distances are all entangled. Figure \ref{fig1:subfigure-b} shows that after injecting energy into one of the systems and making a measurement, the remaining subsystem remains thoroughly entangled.

\section{Circuit Implementation}
\begin{figure}[p]
    
        \centering
        \includegraphics[width=\columnwidth,valign=t]{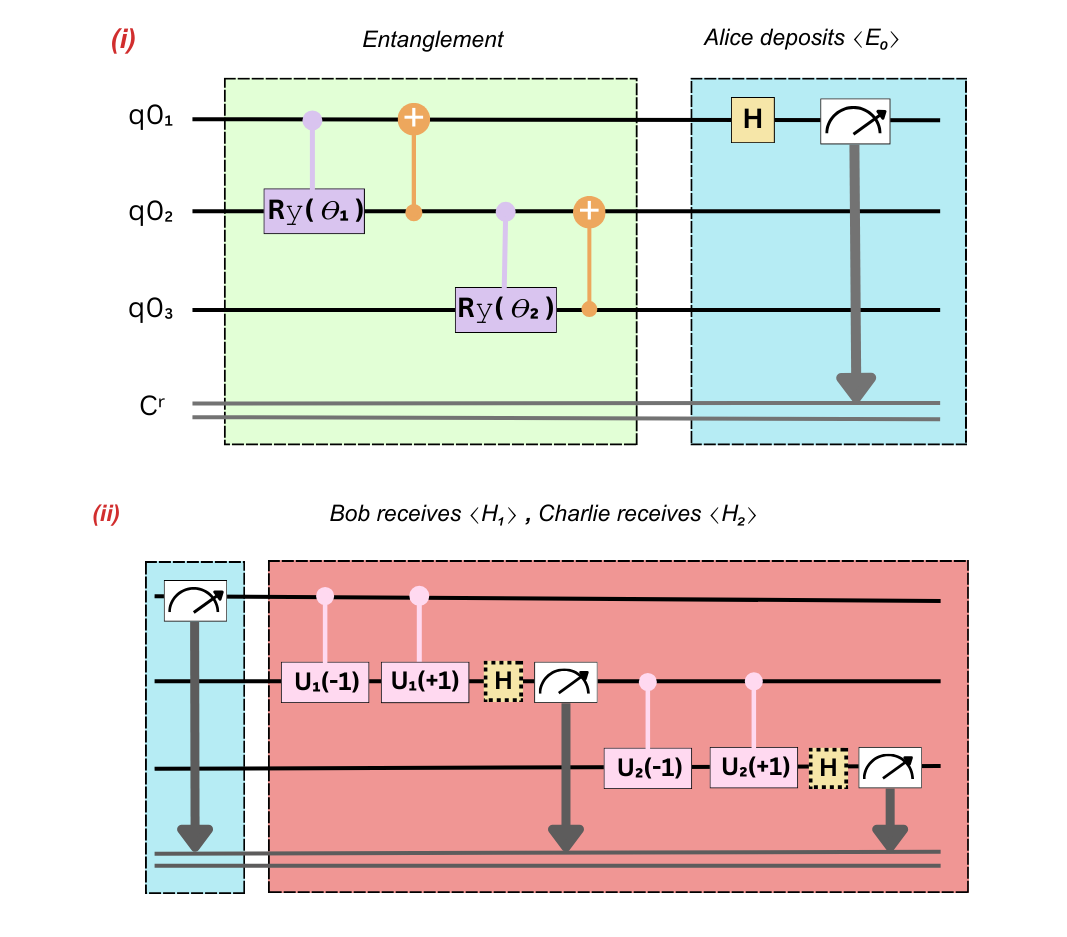}
        \caption{3 qubit systems for quantum energy teleportation. Here, (i) represents the circuit portion where the subsystems are entangled. Immediately after entanglement, Alice deposits energy into the system through projective measurement. (ii) shows the part of the circuits where the measurement outcome is shared with the receivers, Bob and Charlie, via a classical channel.}
       \label{fig2a}
    \end{figure}
\begin{figure}[p]
        \centering
        \includegraphics[width=\columnwidth,valign=t]{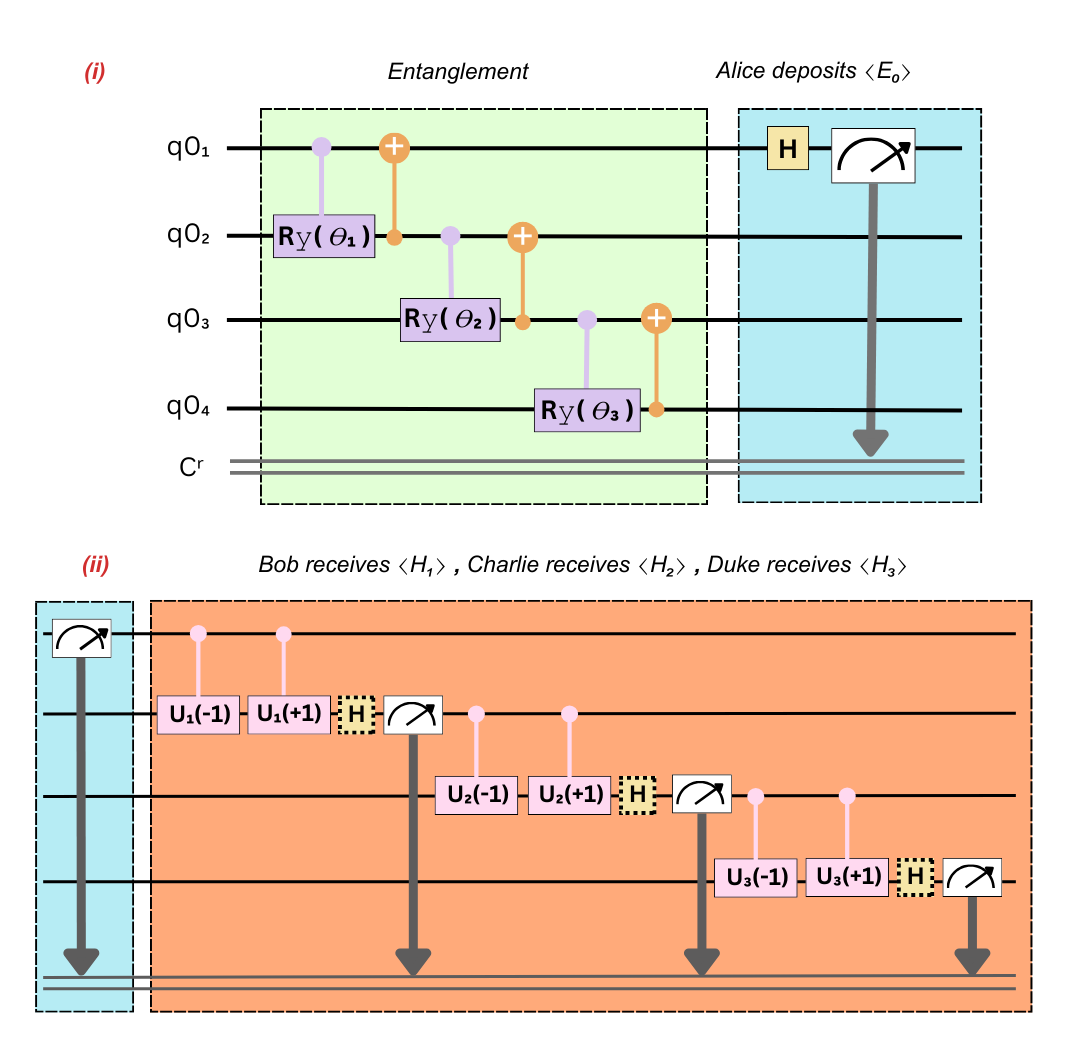}
        \caption{4 qubit system for quantum energy teleportation. Here, (i) represents the circuit portion where the subsystems are entangled. Immediately after entanglement, Alice deposits energy into the system through projective measurement. (ii) shows the part of the circuits where the measurement outcome is shared with the receivers, Bob, Charlie, and Duke via a classical channel.}
        \label{fig2b}
    \end{figure}
    \begin{figure}[p]

        \centering
        \includegraphics[width=\columnwidth,valign=t]{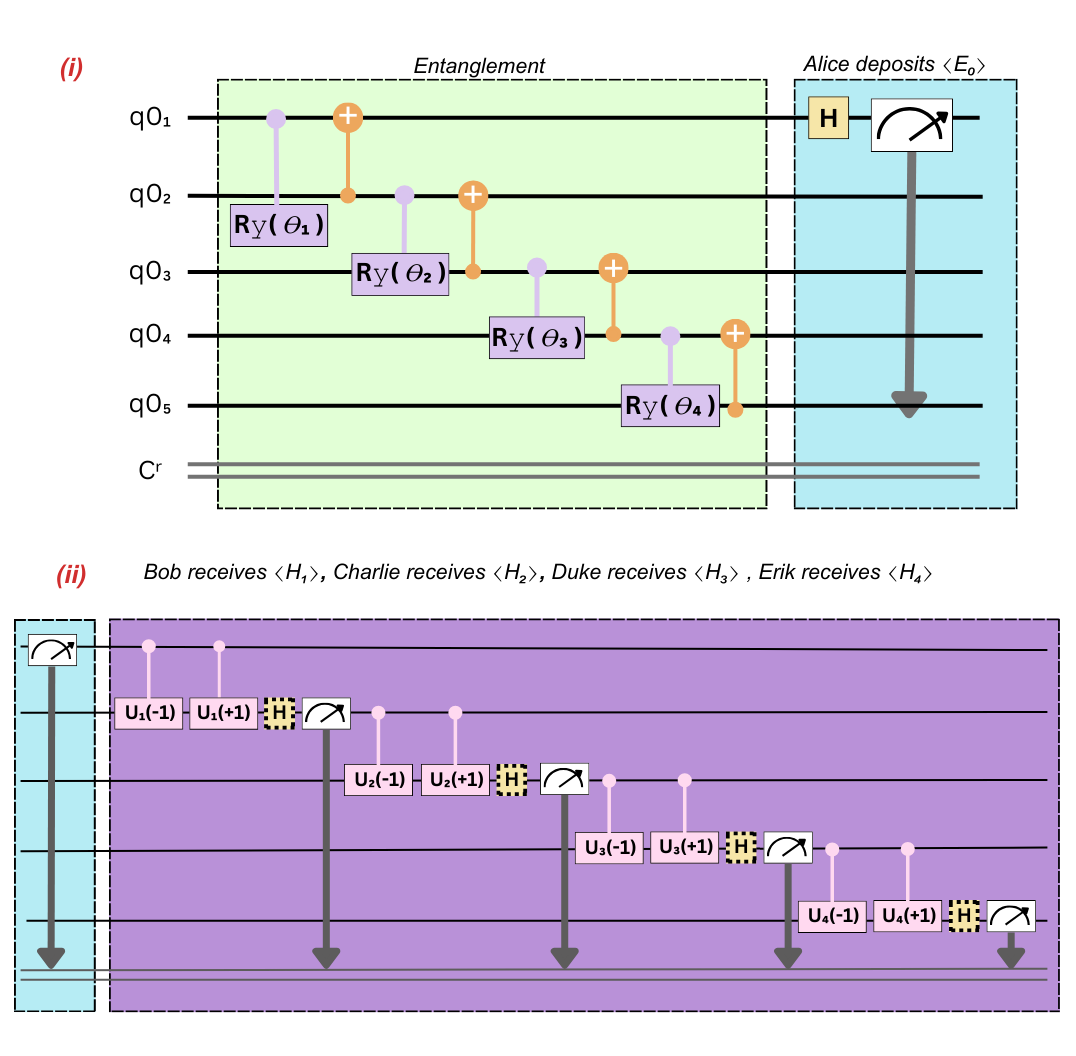}
        
    
    \caption{5 qubit system for quantum energy teleportation. Here, (i) represents the circuit portion where the subsystems are entangled. Immediately after entanglement, Alice deposits energy into the system through projective measurement. (ii) shows the part of the circuits where the measurement outcome is shared with the receivers, Bob, Charlie, Duke, and Erik, via a classical channel.}
   \label{fig2c}
\end{figure}

We have implemented several multi-qubit QET circuits incorporating multiqubit QET protocols as shown in figure \ref{fig2a}, \ref{fig2b} and \ref{fig2c}. The circuit in figure \ref{fig2a} uses three qubits where q0 is representative of Alice (the sender) while q1 and q2 are representative of Bob and Charlie, respectively (the receivers). At first, the particles are entangled using W-state entanglement. Then Alice deposits energy into her system and sends the information about her measured state to Bob or Charlie. Bob and Charlie can measure the energy in their subnetwork by passing the information sent by Alice to their unitary operators. Similarly, the circuit is extended for additional qubits using the same protocol. Figure \ref{fig2b} shows the circuit connection for four qubits, and figure \ref{fig2c} is for five qubits.

Our proposed architecture uses a reduced number of gates compared to the traditional implementation of W-state entanglement \cite{cruz2019efficient}. In doing so, it reduces the time complexity from linear $O(N)$ to logarithmic $O(log_{2}N)$. So, our circuit is fast and optimized.

\subsection{Preparing Entanglement}

GHZ and W-state are classic representations of the general multipartite entanglement \cite{coecke2010compositional}. While GHZ is more fragile against loss, W-state is more robust against it \cite{dhondt2006computational}. W representation is central to quantum memories, multiparty quantum network protocols, and universal quantum cloning machines. 

Another critical aspect of entanglement is that if one state is measured, all other states collapse. However, each receiver should be able to measure locally without collapsing the entanglement between other pairs. So, we used a W-state entanglement between each pair to ensure that the measurement between one does not affect the other. 

W-state \cite{wang2008preparation} can be generalized for n qubits as shown in equation \ref{eq1}. An example of three qubits is mentioned in equation \ref{3q}. Here, the quantum superposition exists with equal expansion coefficients for all possible pure states in which precisely one of the qubits is in an ``excited state," whereas all others are in the ``ground state."

\begin{equation}
\label{eq1}
    \ket{W_{n}} ={\frac {1}{\sqrt {n}}}(|100...0\rangle +|010...0\rangle +...+|00...01\rangle )
\end{equation}

\begin{equation}
\label{3q}
    \ket{W_{3}} ={\frac {1}{\sqrt {3}}}(|100\rangle +|010\rangle +|001\rangle )
\end{equation}

The robustness against particle loss and the LOCC-inequivalence with the (generalized) GHZ state also hold for the n-qubit W state. We have chosen the n-qubit W-state entanglement for our architecture to incorporate these advantages. 

In this entanglement scenario, if the ground state is separable, the subsystems do not correlate. So, to teleport energy, we bring the systems to their ground states and create some ground-state entanglement \cite{hotta2010ground}. Then, the entire system has zero-point energy. The total Hamiltonian of the system can be described by equation \ref{eq_total_ham}, where the Hamiltonians of each local system sum up to the total Hamiltonian.

\begin{equation}\label{eq_total_ham}
  H_{\text{total}} \;=\; \sum_{i=0}^{N-1} H_i \;+\; V 
\end{equation}

Each Hamiltonian may have its own substructure or components hosting multiple subsystems:

\begin{equation}
  H_{\text{total}} \;=\; \sum_{i=0}^{m-1} H_i \;+\; H_{sub_m} \;+\; V ,
\end{equation}

\begin{equation}
  H_{sub_m} \;=\; \sum_{i=m}^{N-1} H_i 
\end{equation}

\begin{figure}[htbp]
    \centering
    \includegraphics[width=0.465\textwidth,valign=t]{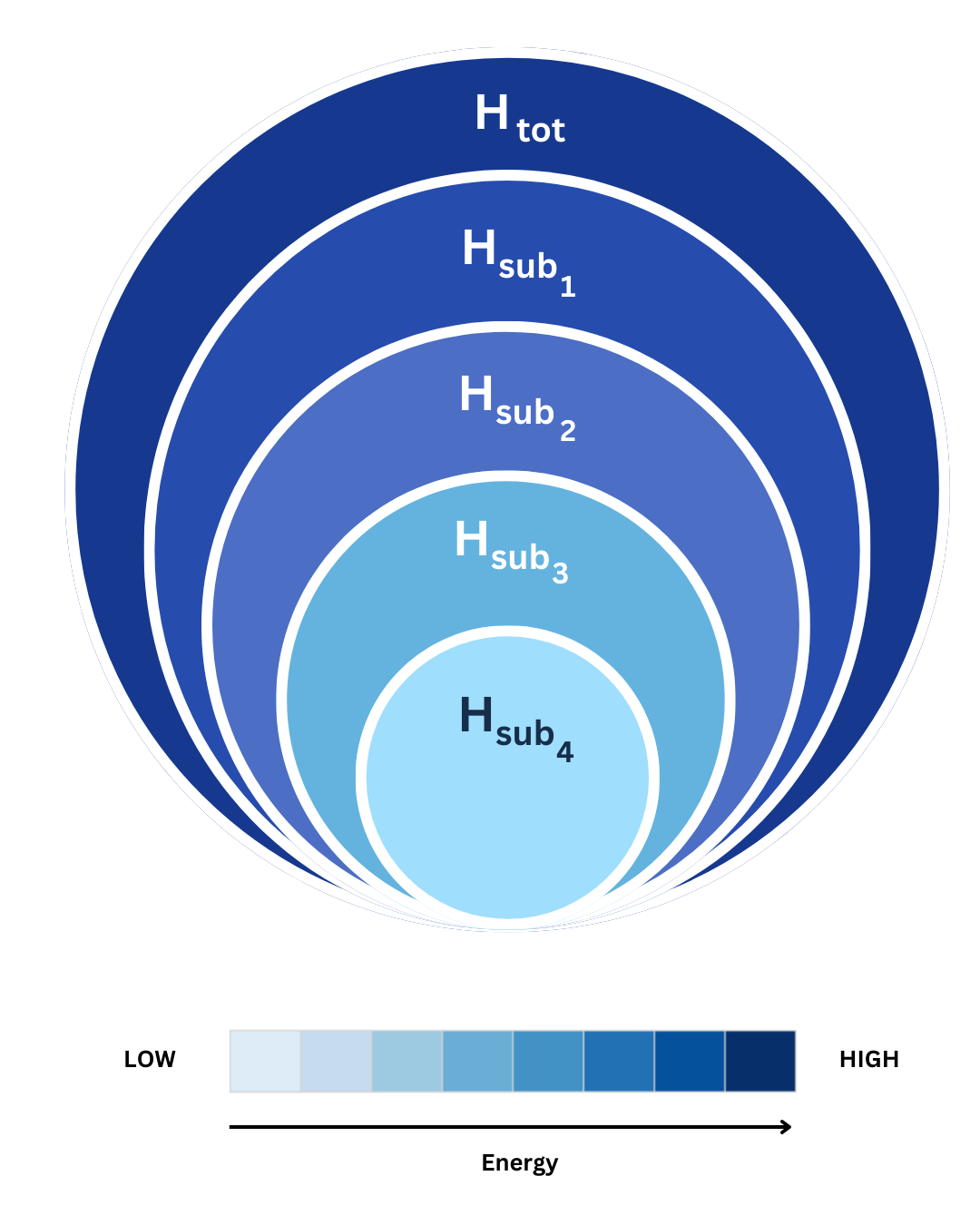}
    \caption{Total Hamiltonian includes Hamiltonian of each subsystem. Again, each subsystem may also have substructures with Hamiltonians that sum up to the subsystem's total Hamiltonian.}
    \label{fig:energy_venn}
\end{figure}

\subsection{Injecting Energy}

Multiqubit QET is carried out in a quantum many-body system. It is known that any non-trivial local operations performed on a quantum many-body system give rise to excited states \cite{liu2023probing}. This includes measurements of the many-body's ground state, increasing the energy expectation value. So, in our experiment, the energy is injected into the sender's system by the experimental devices used for measurement. This projective measurement occurs with the following operation where $\mu$ is the outcome.
\begin{equation}
    \sigma(\mu)= \frac{1}{2}(1+\mu X_{0}).
\end{equation}

Since the quantum many-body system is entangled \cite{amico2008entanglement}, local injection of energy affects the global ground states, i.e., energy is supplied to the entire system \cite{hotta2010energy}. However, the local injection occurred through measurement, destroying the entanglement between the two subsystems. The information about the injected energy is passed on to the entire system before this because the subsystems were already entangled. The measured outcome of the sender contains the information necessary to extract the energy from the subsystems. 

After the $W$–state entangling gates are applied, the $N$–qubit register is
brought to the \emph{normalised} superposition  
\begin{equation}
\label{eq:init-state}
   |\psi_{\text{init}}\rangle
   \;=\;
   \frac{k\,|0\!\cdots\!0\rangle \;+\; h\,|W_{N}\rangle}
        {\sqrt{h^{2}+k^{2}}},
   \qquad h,k\in\mathbb{R}_{\ge0}.
\end{equation}
The real parameters \(h\) and \(k\) tune the relative weight of the true
vacuum and the single-excitation $W_{N}$ component; they completely
determine the energy that can be unlocked in the subsequent QET protocol.

The expectation value of the energy injected into the system is: 
\begin{equation}
  E_{0}\;=\;\frac{h^{2}}{{h^{2}+k^{2}}}.
\end{equation}

Before injecting the energy, the entire system is brought to ground-point energy. At that point, the energy injected into the system is equivalent to the total Hamiltonian of the system. 

\begin{equation}
H_{total} = E_0
\end{equation}

The entire system gains energy by injecting energy into one of the participants. The individual Hamiltonians incorporate this information.

\begin{equation}
H_{n}\;=\;\ket{1}_{n}\!\bra{1}
        \;=\;\frac{1}{2}\bigl(I - Z_{n}\bigr),
\qquad n=0,\dots ,N-1,
\end{equation}

\subsection{Classical Communication}

Once the sender makes a measurement, the entanglement breaks, and there is a measurement outcome of either $\mu$   = $- 1$ or $+1$. This state information is then passed via a classical channel to the receivers, who can use this information to harvest the energy from their local systems. This information is sent as classical bits, and without this information, the receivers cannot obtain the target state. Because classical information needs to be sent, quantum teleportation cannot occur faster than the speed of light. 
After receiving the classical bit $\mu=\pm1$,
each receiver applies the conditional operation  
\[
   U_{\mu} \;=\;
   \begin{cases}
     I, & \mu = +1,\\[6pt]
     Z, & \mu = -1,
   \end{cases}
\]
on its own qubit.
This implements $\sigma_{\mu}^{-1}$, erases the $\mu$-dependent phase,
and converts the remote negative-energy pocket into a positive local
excitation ready for harvesting.  

\subsection{Harvesting Energy}

The post-measurement state $\sigma_{\mu}\ket{\psi}$ of the receiver's subsystem is different from its pre-measurement state $\ket{\psi}$. The receiver can only know this post-measurement state through the classical channel. Once Alice observes the state $ \mu \in \{-1, 1\}$ using local measurement, she sends it to the receivers. When the information reaches the receivers, they can correct the state by performing a unitary operation $\sigma_{\mu}^{-1}$ on their subsystem. Then, the receivers are ready to perform local measurements on their subsystems to harvest the energy injected through entanglement in their ground states.

After measuring Alice's local energy, the entire system can be reimagined in terms of two subsystems where Alice is separated from the remaining entangled structure. 

\begin{equation}
H_{total} = H_0 + H_{sub1} + V
\end{equation}
where
\begin{equation}
H_{sub1} = H_1 + H_{sub2}
\end{equation}

Similarly, after measuring Bob:

\begin{equation}
H_{total} = H_0 + H_1 + H_{sub_2} + V
\end{equation}

Finally, for a three-qubit system, after measuring Charlie:

\begin{equation}
H_{total} = H_0 + H_1 + H_2 + V
\end{equation}

\noindent
Because all qubits are now in product states, the entangling interaction
has vanished,
\[
   V \;\longrightarrow\; 0,
\]
so that the final Hamiltonian is simply
$H_{\text{total}} = \sum_{i=0}^{N-1} H_{i}$
with no residual coupling.  

\section{Symmetry Test}

\subsection{Translational Symmetry}

\begin{figure}[!h]
    \centering
        \includegraphics[width= 0.61\textwidth,valign=t]{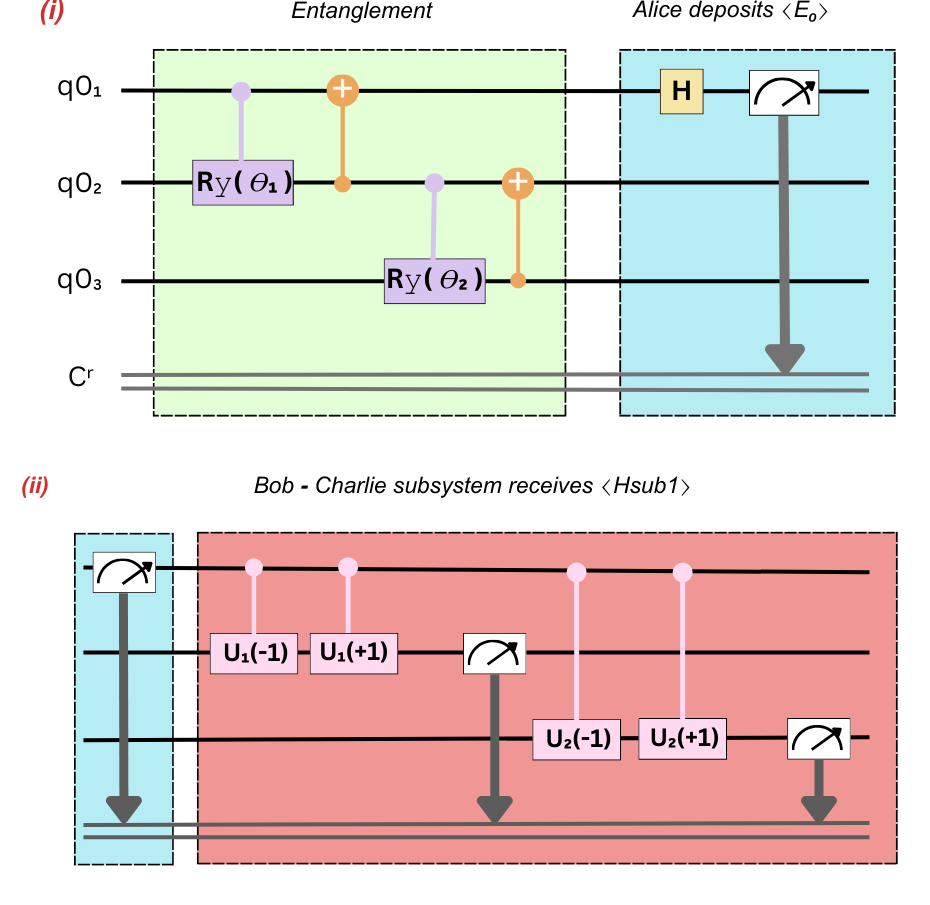}
        \caption{Translational Symmetry: The Hamiltonian of the subsystem measured from any of its nodes gives the same result.}
        \label{sym1}
   \end{figure}
   
Translational symmetry implies that a system's laws of physics, properties, and behavior remain unchanged under spatial translations or shifts. In simpler terms, it means that the physical properties of a system do not depend on its position in space. If we measure the total Hamiltonian of a subsystem, then it must be the same from any subsystem node. Figure \ref{sym1} shows our circuit design for testing this. After measuring Alice's energy, the remaining subsystem has a Hamiltonian of $H_{sub_1}$, which is measured to be the same from any node of the subsystem. So, our design passes the test.

\subsection{Exchange Symmetry}

\begin{figure}[!h]
       \centering\includegraphics[width=0.61\textwidth,valign=t]{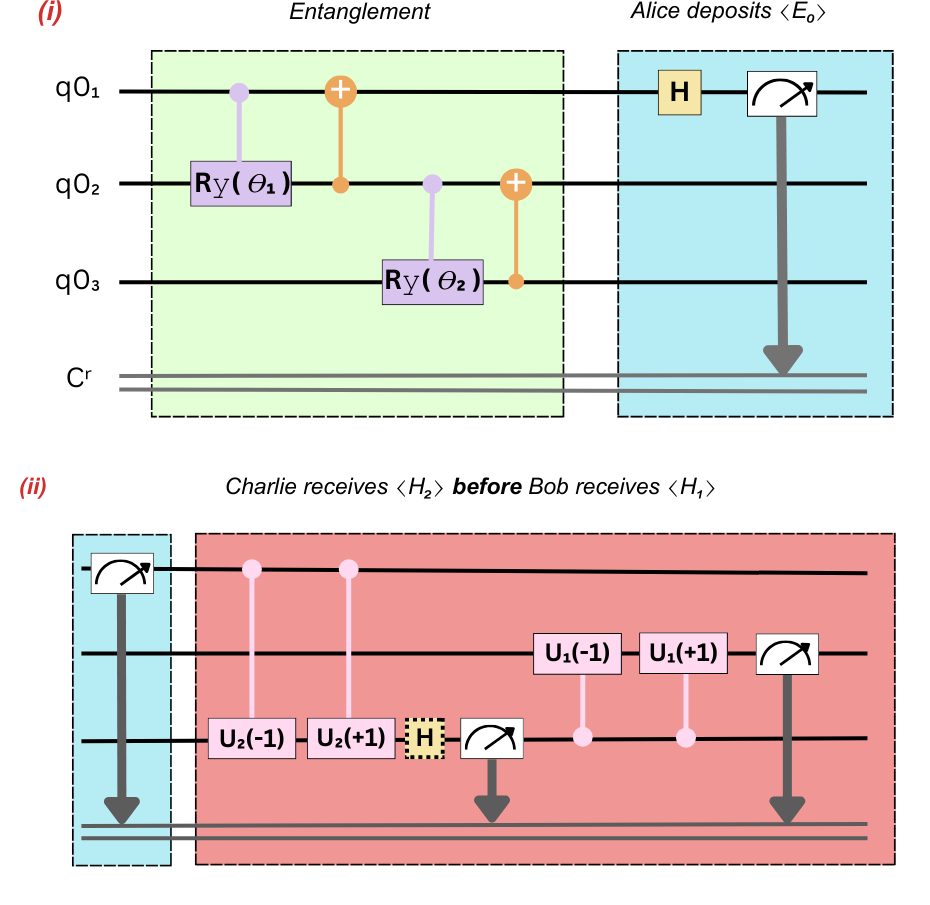}
     \caption{Exchange Symmetry: Charlie and Bob switch order of measurement.}
        \label{sym2}
\end{figure}

Exchange symmetry refers to the principle that identical particles cannot be distinguished from one another. If you have two identical particles and exchange their positions in a system, the system's properties should remain unchanged. 
Figure \ref{sym2} shows the circuit design where the third qubit (Charlie) is measured before the second qubit (Bob). The measurement outcomes remain unchanged even after exchanging the order.
\begin{equation}
\psi(A,B) = \psi(B,A)
\end{equation}

\section{Results Analysis}
\begin{figure}[!htb]
    \centering
    \begin{subfigure}{0.7\textwidth}
        \centering
        \includegraphics[width=\linewidth,valign=t]{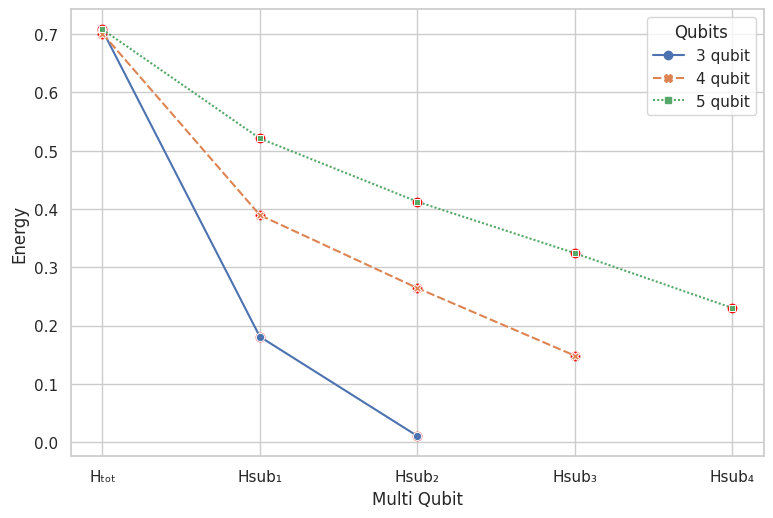}
        \caption{Simulation(qasm-simulator)}
        \label{fig4:subfigure-a}
    \end{subfigure}
    \hfill
    \begin{subfigure}{0.7\textwidth}
        \centering
       \includegraphics[width=\columnwidth,valign=t]{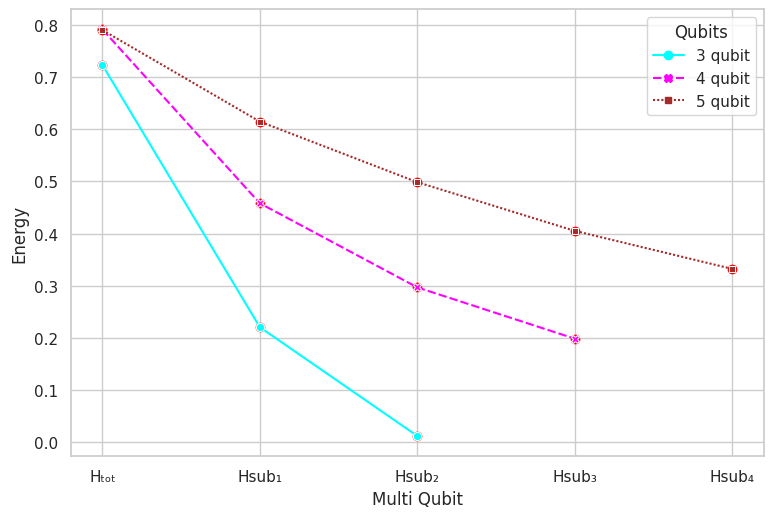}
     \caption{Real Quantum Device (ibm\_lagos)}
        \label{fig4:subfigure-b}
    \end{subfigure}
    
   \caption{The energy readings from each subsystem in the multi-qubit system show a decremental energy distribution.} 
     \label{figcomp}  
\end{figure}

The energy readings obtained from our QET experiment across multi-qubit systems unveil some exciting patterns. For a three-qubit system, Table \ref{tab3} shows that the energy readings obtained by measuring the qubits gradually decrease according to who measures first. This is because each Hamiltonian incorporates the Hamiltonians of their substructures. Similarly, Table \ref{tab4} and \ref{tab5} demonstrate 4 and 5  qubit systems. The graphical representation of the combined values from Table \ref{tab3}, \ref{tab4}, and \ref{tab5} is seen in Figure \ref{figcomp}. Both Figure \ref{fig4:subfigure-a} and \ref{fig4:subfigure-b} show the gradual decrement of the Hamiltonians of the subsystems for qasm\_simulator and real device, respectively. It is essential to understand that the readings are not for a particular qubit but for the entangled substructure. These substructures contain several qubits with individual Hamiltonians. So, the energy the sender initially injects is first distributed into the entire system. Then, the share of individual qubits is learned through measurement. The sum of the individual qubit energies never exceeds the amount of injected energy. There is conservation of energy. Some energy injected into the system is lost, and the remaining is dispersed throughout the system. We recorded the energy measurements in both simulation and real quantum devices. In both platforms, the same relation exists where the injected energy is distributed from the sender to the entire system, and the subsystems contain the energy in a decremental order followed by measurement. This is because in W-state entanglement, even after the measurement of a qubit, only the entanglement between that qubit and others is reduced to a single state, whereas all other qubits remain entangled. So, the remaining entangled subsystem has energy dispersed within its network. The same scenario exists across all of the multiqubit systems as shown in Table \ref{tab3}, \ref{tab4}, \ref{tab5}, and figure \ref{figcomp}. 

\section{Real Quantum Device}
\begin{figure}
    \centering
    
    \begin{subfigure}{0.8\textwidth}
        \centering
        \includegraphics[width=1\columnwidth,valign=t]{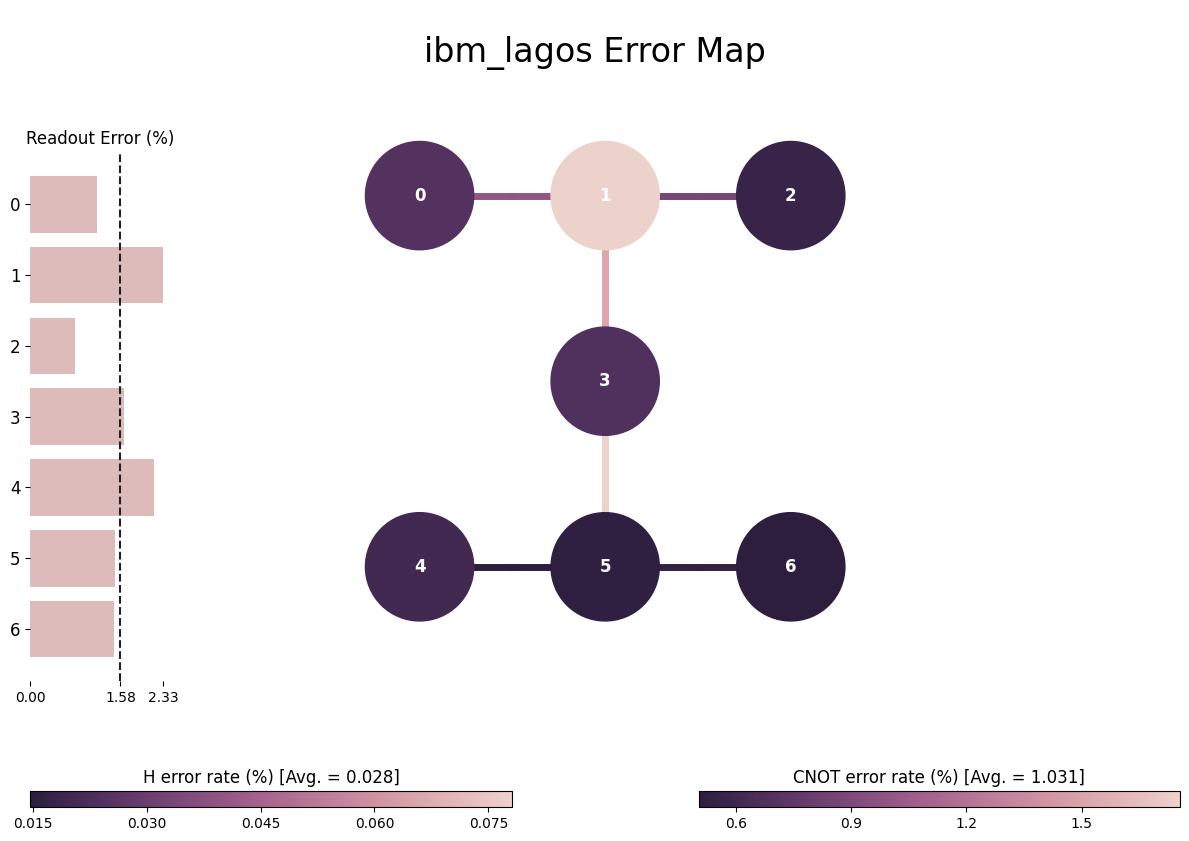}
        \caption{ibm\_lagos}
        \label{errorfigl}
 \end{subfigure}

   \caption{Error Map }
        \label{figerror}
     \end{figure}

Our study was not limited to simulation only. We have run extensive experimentation on real quantum computers provided by IBM, which anyone can access. The device available to us was IBMQ Lagos. Figure \ref{figerror} shows the error maps of these devices, which represent the error rates associated with the various quantum gates and operations performed on the quantum processor. It is crucial to characterize and understand the errors that occur during the execution of quantum circuits. This is because these errors can limit the performance and reliability of quantum algorithms. Despite these challenges, real quantum devices provide valuable insights into the practical limitations and opportunities for quantum computing, and simulations remain important tools for algorithm development and benchmarking. The comparisons between the simulation values and real values in Table \ref{tab3}, \ref{tab4}, and \ref{tab5} shed some light on how real device noise makes the readings slightly different from the simulation. However, due to the simplified architecture used in this study, we observed a noticeably low error deviation rate even with a higher circuit depth.

Our setup has no significant deviation of results from real quantum devices, so our multiqubit energy teleportation system is feasible for real-world applications. The proposed multiqubit QET protocol stands robust against real-world quantum device limitations. With additional error-mitigation strategies the accuracy could be further improved \cite{takagi2022fundamental}.

\section{Conclusion}

We have presented the first experimental realisation of \emph{multi-qubit} quantum-energy teleportation (QET) using a robust $W$-state architecture. Our three-, four-, and five-qubit circuits were executed both in noiseless simulation and on the \texttt{IBMQ Lagos} superconducting processor, demonstrating that a single sender can inject an energy \(E_{0}\) that is then \emph{deterministically and decrementally} harvested by multiple, spatially separated receivers.  
Energy conservation and causality are respected throughout: the sum of all harvested energies never exceeds \(E_{0}\), and classical feed-forward limits the protocol to light-speed signalling.

Two internal consistency checks—translational symmetry of the residual sub-Hamiltonian and exchange symmetry under permutation of receiver order—were passed on both the simulator and hardware runs, confirming that the observed distribution pattern is an intrinsic feature of the $W$-state QET protocol.  

The present proof-of-principle is limited by NISQ-era (noisy intermediate-scale quantum) noise and by the single-excitation nature of the $W$ state, so only a fraction of the injected energy is ultimately recovered. Nevertheless, today’s quantum
hardware is already powerful enough to implement and verify the protocol up to five qubits, and error-mitigation techniques promise even higher fidelity.

Looking forward, larger entangled networks, alternative resource states, and hybrid photonic-superconducting platforms could enable long-range energy distribution  or even probing topological order via QET spectroscopy. Our results therefore open a practical path toward energy-aware quantum networks and deepen the link between entanglement, information and thermodynamics.

\begin{table}[!htb]
\centering
\caption{Energy readings from QET experiment for three qubits. $E_0$ is the energy injected, $H_n$ is the hamiltonian of the nth qubit and $H_{sub_n}$ is the hamiltonian of the nth subsystem where $H_{sub_m}=\displaystyle\sum_{i=m}^{N-1} H_{i}\,$.} 
\label{tab3}
\large
\renewcommand{\arraystretch}{1.3} 
 \begin{tabularx}{\columnwidth}{X l l X}
\toprule
\textbf{Energy} & \textbf{h, k} & \textbf{qasm simulator} & \textbf{$IBMQ Lagos$}  \\
\midrule

\multirow{2}{*}{$H_{tot}$}  & {2,1} & 1.7845 ± 0.0063 & 1.8627 ± 0.0063 \\
                      & {1,1} &  0.7064 ±  0.0031 & 0.7321 ± 0.0031 \\
                    
\midrule
\multirow{2}{*}{$H_{sub_1}$} & {2,1} & 0.7314 ± 0.0053 & 0.9611 ± 0.0057\\
                      & {1,1} & 0.1809  ±  0.0026 & 0.2206 ± 0.0028\\
                     
\midrule
\multirow{2}{*}{$H_{sub_2}$} & {2,1} & 0.3918  ± 0.0045 & 0.4675± 0.0058  \\
                      & {1,1} & 0.0107  ±  0.0022 & 0.0122 ± 0.0028\\
                
\midrule
\multirow{2}{*}{$E_{o}$}  & {2,1} & 1.7888 & 1.7888 \\
                      & {1,1} &  0.7071 & 0.7071 \\
\midrule
\multirow{2}{*}{$H_{1}$}  & {2,1} & 1.0531 & 0.9016 \\
                      & {1,1} & 0.5255 & 0.5115\\
                     
\midrule
\multirow{2}{*}{$H_{2}$}  & {2,1} & 0.3396 & 0.4936 \\
                      & {1,1} &  0.1702 & 0.2084 \\                 
\bottomrule
\end{tabularx}
\end{table}

\begin{table}[!htb]
\centering
\caption{Energy readings from QET experiment for four qubits. $E_0$ is the energy injected, $H_n$ is the hamiltonian of the nth qubit and $H_{sub_n}$ is the hamiltonian of the nth subsystem where $H_{sub_m}=\displaystyle\sum_{i=m}^{N-1} H_{i}\,$.} 

\label{tab4}
\large
\renewcommand{\arraystretch}{1.3} 
 \begin{tabularx}{\columnwidth}{X l l X}
\toprule
\textbf{Energy} & \textbf{h, k} & \textbf{qasm simulator} & \textbf{$IBMQ Lagos$}  \\
\midrule

\multirow{2}{*}{$H_{tot}$}  & {2,1} & 1.7846 ± 0.0063 & 1.9186 ± 0.0063\\
                      & {1,1} &  0.7009 ± 0.0031 & 0.7915 ± 0.0031\\
                    
\midrule
\multirow{2}{*}{$H_{sub_{1}}$} & {2,1} & 1.1575 ± 0.0060 & 1.4426 ± 0.0062 \\
                      & {1,1} & 0.3895 ± 0.0029 & 0.4582 ± 0.0031 \\
                     
\midrule
\multirow{2}{*}{$H_{sub_{2}}$} & {2,1} & 0.9156 ± 0.0056 & 1.0287 ± 0.0061\\
                      & {1,1} & 0.2644 ± 0.0028 & 0.2970 ± 0.0030\\
                      
\midrule
\multirow{2}{*}{$H_{sub_{3}}$} & {2,1} & 0.6624 ± 0.0052 & 0.7885 ± 0.0060 \\
                      & {1,1} & 0.1485 ± 0.0026 & 0.1987 ± 0.0031\\

\midrule
\multirow{2}{*}{$E_{o}$}  & {2,1} & 1.7888 & 1.7888 \\
                      & {1,1} &  0.7071 & 0.7071 \\
\midrule
\multirow{2}{*}{$H_{1}$}  & {2,1} & 0.6271 & 0.476\\
                      & {1,1} & 0.3114 & 0.3333\\
                     
\midrule
\multirow{2}{*}{{$H_{2}$}}  & {2,1} & 0.2419 & 0.4139 \\
                      & {1,1} & 0.1251 & 0.1612 \\
                     
\midrule
\multirow{2}{*}{$H_{3}$}   & {2,1} & 0.2532 & 0.2402 \\
                      & {1,1} & 0.1157 & 0.0983 \\

\bottomrule
\end{tabularx}
\end{table}

\begin{table}[!htb]
\centering
\caption{{Energy readings from QET experiment for five qubits. $E_0$ is the energy injected, $H_n$ is the hamiltonian of the nth qubit and $H_{sub_n}$ is the hamiltonian of the nth subsystem where $H_{sub_m}=\displaystyle\sum_{i=m}^{N-1} H_{i}\,$.} }

\label{tab5}
\large
\renewcommand{\arraystretch}{1.1} 
\begin{tabularx}{\columnwidth}{X l l X}
\toprule
\textbf{Energy} & \textbf{h, k} & \textbf{qasm simulator} & \textbf{$IBMQ Lagos$}  \\
\midrule
\multirow{2}{*}{$H_{tot}$}  & {2,1} & 1.7830 ± 0.0063 & 1.9488 ± 0.0063\\
                      & {1,1} & 0.7079  ± 0.0031 & 0.7902 ± 0.0031\\
                    
\midrule
\multirow{2}{*}{$H_{sub_{1}}$} & {2,1} & 1.4128 ± 0.0062 & 1.77989 ± 0.0063\\
                      & {1,1} & 0.5210 ± 0.0031 & 0.6148 ± 0.0031 \\
                     
\midrule
\multirow{2}{*}{$H_{sub_{2}}$} & {2,1} & 1.2120 ± 0.0060 & 1.4866 ± 0.0063\\
                      & {1,1} & 0.4125 ± 0.0030 & 0.4987 ± 0.0031\\
                      
\midrule
\multirow{2}{*}{$H_{sub_{3}}$} & {2,1} & 1.0266 ± 0.0058 & 1.2777 ± 0.0063 \\
                      & {1,1} & 0.3243 ± 0.0029 & 0.4054 ± 0.0031\\
\midrule
\multirow{2}{*}{$H_{sub_{4}}$} & {2,1} & 0.8325 ± 0.0055 & 1.0973 ± 0.0063\\
                      & {1,1} & 0.2303 ± 0.0027& 0.3325 ± 0.0031\\
\midrule
\multirow{2}{*}{$E_{o}$}  & {2,1} & 1.7888 & 1.7888 \\
                      & {1,1} &  0.7071 & 0.7071 \\
\midrule
\multirow{2}{*}{$H_{1}$}  & {2,1} & 0.3702 & 0.169 \\
                      & {1,1} & 0.1869 & 0.1754 \\                
\midrule
\multirow{2}{*}{$H_{2}$}  & {2,1} & 0.2008 & 0.2932 \\
                       & {1,1} & 0.1085 & 0.1161 \\             
\midrule
\multirow{2}{*}{$H_{3}$}   & {2,1} & 0.1854 & 0.2089\\
                       & {1,1} & 0.0882 & 0.0933\\
\midrule
\multirow{2}{*}{$H_{4}$} & {2,1} & 0.1941 & 0.1797 \\
                      & {1,1} & 0.094 & 0.0729 \\
\bottomrule
\end{tabularx}
\end{table}
\clearpage

 \bibliographystyle{qet} 
 \bibliography{qet}





\end{document}